\documentstyle[aps,pra,epsf]{revtex}
\begin{document}
\draft

\newcommand{\pp}[1]{\phantom{#1}}
\newcommand{\be}{\begin{eqnarray}}
\newcommand{\ee}{\end{eqnarray}}
\newcommand{\ve}{\varepsilon}
\newcommand{\vs}{\varsigma}
\newcommand{\Tr}{{\,\rm Tr\,}}
\newcommand{\pol}{\frac{1}{2}}

\title{
Noncanonical quantum optics
}
\author{Marek~Czachor}
\address{
Katedra Fizyki Teoretycznej i Metod Matematycznych\\
Politechnika Gda\'{n}ska,
ul. Narutowicza 11/12, 80-952 Gda\'{n}sk, Poland\\
and\\
Arnold Sommerferld Instit\"ut f\"ur Mathematische Physik\\
Technische Universit\"at Clausthal, 38678 Clausthal-Zellerfeld, Germany}
\maketitle

\begin{abstract}
Modification of the right-hand-side of canonical commutation relations (CCR)
naturally occurs if one considers a harmonic oscillator with indefinite 
frequency. Quantization of electromagnetic field by means of such a non-CCR
algebra naturally removes the infinite energy of vacuum but still results 
in a theory which is very similar to quantum electrodynamics. 
An analysis of perturbation theory shows that the non-canonical theory 
has an automatically built-in cut-off but requires charge/mass renormalization
already at the nonrelativistic level. 
A simple rule allowing to compare perturbative 
predictions of canonical  and non-canonical theories is given.  
The notion of a unique vacuum state is 
replaced by a set of different vacua. Multi-photon states 
are defined in the standard way but depend on the choice of vacuum. 
Making a simplified choice of the vacuum state we estimate corrections
to atomic lifetimes, probabilities of multiphoton 
spontaneous and stimulated emission, and the Planck law. 
The results are practically identical to the standard ones. Two different 
candidates for a free-field Hamiltonian are compared.
\end{abstract}

%\narrowtext

\section{Introduction}

The standard quantization of a harmonic oscillator is based on 
quantization of $p$ and $q$ but $\omega$ is a parameter. To have,
say, two different frequencies one has to consider two independent 
oscillators. On the other hand, it is evident that there can exist
oscillators which are in a {\it quantum superposition\/} of different
frequencies. The example is an oscillator wave packet associated with
distribution of center-of-mass momenta. It is known that the superposition of 
momenta gets translated into a superposition od Doppler shifts 
and therefore also of frequencies. We stress here the word ``quantum"
since the superpositions we have in mind are not those we know from
{\it classical\/} oscillations. 

This trivial observation raises the question of the role of
superpositions of frequencies for a description of a single harmonic
oscillator. The motivation behind the problem is associated with the question
of field quantization: Is it possible that a quantum field consists 
of oscillators whose frequencies are {\it indefinite\/}? If so, maybe to
quantize the field it is sufficient to use only one oscillator which exists 
in a {\it quantum\/} superposition of all the possible frequencies allowed 
by the boundary conditions of a given problem? 

The idea is very simple. It is known that a ``one-particle'' 
state vector can be regarded as a representation of an ensemble of particles 
in a given pure state. On the other hand, the classical electromagnetic 
field can be regarded as an ensemble of oscillators. The standard idea of 
quantization, going back to 1925 \cite{BHJ}, 
is to treat the field as an ensemble of 
quantum oscillators. But the ensemble itself is, in a sense, a classical one 
since for each frequency we need a separate oscillator. This is analogous 
to a classical ensemble of particles forming a classical wave on a lake 
surface.
For each point on the surface we need a separate particle because a
classical particle can ocupy only a single point in space. A quantum wave 
is of course different and we are all accustomed to the idea of a 
single-particle wave. In this case the properties of the entire ensemble are 
somehow encoded in properties of a single element of the ensemble. 

For some reasons, probably partly historical, it seems 
that the idea of 
a single-particle state vector representation of the ensemble of oscillators
has never been considered. 
The historical reason may be the fact that the very concept of field 
quantization occured already in 1925. At that stage quantum mechanics existed 
still in a matrix form and the Schr\"odinger paper ``Quantisierung als 
Eigenwertproblem''\cite{Sch}, where the Schr\"odinger equation occured for the
first time and the role of eigenvalues was explained, was not yet published. 
Actually, as explained by Jammer \cite{Jammer}, Heisenberg quantized the harmonic 
oscillator without having heard of matrices so no wonder he could not treat the 
parameter $\omega$ entering $E=p^2/2m+m\omega^2 q^2/2$ as an eigenvalue of some 
operator. 

Heisenberg's quantization leads to the well known algebra of canonical 
commutation relations (CCR) $[a_\omega,a_\omega^{\dag}]=\bbox 1$, where $\bbox 1$ 
is the identity operator and $\omega$ a classical parameter. 
As it turns out the replacemwnt of $\omega$ by an operator $\hat\omega$ leads to 
non-canonical commutation relations (non-CCR)  
$[a_\omega,a_\omega^{\dag}]=\bbox 1_\omega$, where $\omega$ is an eigenvalue of 
$\hat\omega$ and $\sum_\omega \bbox 1_\omega=\bbox 1$. $\bbox 1_\omega$, similarly
to $\bbox 1$ commutes with all creation and annihilation operators and the 
remaining commutators of non-CCR are the same as those of CCR.
This subtle difference of the right-hand sides of CCR and non-CCR immediately explains 
why a non-CCR-quantized electromagnetic field will have vacuum with finite energy.
Since the non-CCR and CCR algebras are so similar to each other it is not surprizing
that the resulting theories are also very similar. 

The main consequence of the non-CCR modification of CCR is different normalization 
of $n$-photon states. For example, 
\be
\sum_\omega \langle 0|a_\omega a_\omega^{\dag}|0\rangle =1
\ee
and therefore $\langle 0|a_\omega a_\omega^{\dag}|0\rangle<1$.
This should be contrasted with the 
CCR result
 $\langle 0|a_\omega a_\omega^{\dag}|0\rangle=1$, and the resulting divergence 
\be
\sum_\omega \langle 0|a_\omega a_\omega^{\dag}|0\rangle =\infty
\ee
which is {\it the\/} source of infinite vacuum energy. 

To end these introductory remarks one should mention 
that several approaches towards an alternative 
description of the electromagnetic field at a fundamental level were 
already proposed 
(e.g. Janes' \cite{Janes} 
neoclassical theory,  stochastic electrodynamics \cite{MS}). 
But the main  idea of all such alternatives was to treat the 
field in classical terms and to associate the observed discreteness of 
emission/absorbtion phenomena with the quantum nature of atoms and not with
the field itself. 

The approach we will discuss in this paper does not belong to this 
tradition, is much more radical and, so to say, goes 
in the opposite direction. We will not 
try to make the field more classical. What we will try to do is to make 
it even {\it more quantum\/}  
by replacing classical parameters with eigenvalues. 

\section{Harmonic oscillator in superposition of frequencies}

We know that  frequency is typically associated with an
eigenvalue of some Hamiltonian or, which is basically the same, with
boundary conditions. A natural way of incorporating different
frequencies into a single harmonic oscillator is by means of the
{\it frequency operator\/} 
\be
\Omega=\sum_{\omega_k,j_k}\omega_k|\omega_k,j_k\rangle
\langle \omega_k,j_k|
\ee
where all $\omega_k\geq 0$. 
For simplicity we have limited the discussion to the discrete
spectrum but it is useful to include from the outset the possibility
of degeneracies, represented here by the additional discrete quantum
numbers $j_k$. The corresponding Hamiltonian is defined by
\be
H
&=&
\hbar\Omega\otimes \frac{1}{2}\big(a^{\dag}a+aa^{\dag}\big)
\ee
where $a=\sum_{n=0}^\infty\sqrt{n+1}|n\rangle\langle n+1|$. 
The eigenstates of $H$ are $|\omega_k,j_k,n\rangle$ and satisfy the required 
formula
\be
H|\omega_k,j_k,n\rangle= \hbar\omega_k\Big(n+\frac{1}{2}\Big)
|\omega_k,j_k,n\rangle
\ee
justifying our choice of $H$. 
The standard case of the oscillator whose frequency is just $\omega$ 
coresponds either to $\Omega=\omega\bbox 1$ or to the subspace
spanned by $|\omega_k,j_k,n\rangle$ with fixed $\omega_k=\omega$. 
Introducing the operators 
\be
a_{\omega_k,j_k}=|\omega_k,j_k\rangle\langle \omega_k,j_k|\otimes a
\ee
we find that 
\be 
H&=&
\frac{1}{2}\sum_{\omega_k,j_k}\hbar\omega_k
\Big(a_{\omega_k,j_k}^{\dag}a_{\omega_k,j_k} 
+
a_{\omega_k,j_k}a_{\omega_k,j_k}^{\dag}\Big).
\ee
The algebra of the oscillator is ``noncanonical":
\be
{[a_{\omega_k,j_k},a_{\omega_k,j_k}^{\dag}]}&=&
|\omega_k,j_k\rangle\langle\omega_k,j_k|\otimes \bbox 1
=1_{\omega_k,j_k}\\
{[a_{\omega_k,j_k},a_{\omega_l,j_l}^{\dag}]}&=&0
\quad {\rm for}\,(\omega_k,j_k)\neq (\omega_l,j_l)\\
{[a_{\omega_k,j_k},a_{\omega_l,j_l}]}&=& 0\\
{[a_{\omega_k,j_k}^{\dag},a_{\omega_l,j_l}^{\dag}]}&=& 0
\ee
The dynamics in the Schr\"odinger picture is given by 
\be
i\hbar\partial_t|\Psi\rangle
&=&
H |\Psi\rangle=
\hbar\Omega\otimes \big(a^{\dag}a+\frac{1}{2}\bbox 1\big)
|\Psi\rangle.
\ee
In the Heisenberg picture we obtain the important formula 
\be
a_{\omega_k,j_k}(t)
&=&
e^{iHt/\hbar}a_{\omega_k,j_k}e^{-iHt/\hbar}\\
&=&
|\omega_k,j_k\rangle\langle \omega_k,j_k|
\otimes
e^{-i\omega_k t} a
=
e^{-i\omega_k t} a_{\omega_k,j_k}.
\label{exp}
\ee
Taking a general state
\be
|\psi\rangle=\sum_{\omega_k,j_k,n}\psi(\omega_k,j_k,n)|\omega_k,j_k,n\rangle
\ee
we find that the average energy of the oscillator 
is 
\be
\langle H\rangle=
\langle\psi|H|\psi\rangle
=
\sum_{\omega_k,j_k,n}|\psi(\omega_k,j_k,n)|^2
\hbar\omega_k\Big(n+\frac{1}{2}\Big).
\ee
The average clearly looks as an average energy of an {\it ensemble of
different and independent oscillators\/}. The ground state of the
ensemble, i.e. the one with $\psi(\omega_k,j_k,n>0)=0$ 
has energy 
\be
\langle H\rangle=\frac{1}{2}\sum_{\omega_k,j_k}|\psi(\omega_k,j_k,0)|^2
\hbar\omega_k
\ee
which is finite if 
\be
\sum_{\omega_k,j_k}\psi(\omega_k,j_k,0)|\omega_k,j_k\rangle
\ee
belongs to the domain of $\Omega$.
The result is not surprising but still quite remarkable if one thinks
of the problem of field quantization. 

The very idea of quantizing the electromagnetic field, as put
forward by Born, Heisenberg, Jordan \cite{BHJ} and Dirac \cite{D},
is based on the observation that the mode decomposition of the
electromagnetic energy is analogous to the energy of an ensemble of
independent harmonic oscillators. In 1925, after the work of
Heisenberg, it was clear what to do: One had to replace each
classical oscillator by a quantum one. But since each oscillator had
a definite frequency, to have an infinite number of different
frequencies one needed an infinite number of oscillators. 
The price one payed for this assumption was the 
infinite energy of the electromagnetic vacuum. 

The infinity is regarded as an ``easy" one since one can get rid of
it by redefining the Hamiltonian and removing the infinite term. 
The result looks correct and many properties typical of a {\it
quantum\/} harmonic oscillator are indeed observed in electromagnetic
field. However, subtraction of infinite terms is in mathematics 
as forbidden as division by zero so to avoid evident absurdities one
is forced to invent various ad hoc regularizations whose only
justification is that otherwise the theory would not work.  
In larger perspective
(say, in cosmology) it is not at all clear that an infinite (or
arbitrarily cut off at the Planck scale) energy of
the vacuum does not lead to contradictions with observational data 
\cite{Lambda}. 
Finally, Dirac himself had never been fully satisfied by the theory
he created. 
As Weinberg put it, Dirac's  ``demand for a completely
finite theory is similar to a host of other aesthetic judgements that
theoretical physicists always need to make" \cite{Dreams}.

The oscillator that can exist in superpositions of different
frequencies is a natural candidate as a starting point for Dirac-type
field quantization. Symbolically, if the Heisenberg quantization is 
$p^2+\omega^2q^2\mapsto \hat p^2+\omega^2\hat q^2$, where $\omega$ is a
parameter, the new scheme is $p^2+\omega^2q^2\mapsto \hat p^2+\hat
\omega^2\hat q^2$, where $\hat \omega$ is an operator. Its spectrum
can be related to boundary conditions imposed on the fields. 

We do not need to remove the ground state energy since in the Hilbert
space of physical states the correction is finite. The question we
have to understand is whether one can obtain the well known quantum
properties of the radiation field by this type of quantization.

\section{``First quantization'' --- One-oscillator 
field operators}

The new quantization will be performed in two steps. In this section
we describe the first step, a kind of first quantization. In next
sections we shall perform an analogue of second quantization which
will lead to the final framework. 

The energy and momentum operators of the field are defined in analogy to $H$
from the previous section
\be
H &=& 
\sum_{s,\vec\kappa}\hbar \omega_{\vec \kappa}
|s,\vec \kappa\rangle \langle s,\vec \kappa|
\otimes \frac{1}{2}\Big(a^{\dag}a+a a^{\dag}\Big)\\
&=& 
\frac{1}{2}\sum_{s,\vec\kappa}\hbar \omega_{\vec \kappa}
\Big(a_{s,\vec\kappa}^{\dag}a_{s,\vec\kappa} 
+a_{s,\vec\kappa} a_{s,\vec\kappa}^{\dag}\Big)\\
\vec P &=& 
\sum_{s,\vec\kappa}\hbar \vec \kappa
|s,\vec \kappa\rangle \langle s,\vec \kappa|
\otimes \frac{1}{2}\Big(a^{\dag}a+a a^{\dag}\Big)\\
&=& 
\frac{1}{2}\sum_{s,\vec\kappa}\hbar \vec \kappa
\Big(a_{s,\vec\kappa}^{\dag}a_{s,\vec\kappa} 
+a_{s,\vec\kappa} a_{s,\vec\kappa}^{\dag}\Big)
\ee
where $s=\pm 1$ corresponds to circular polarizations. Denote 
$P=(H/c,\vec P)$ and $P\cdot x=Ht-\vec P\cdot \vec x$.
We employ the standard Dirac-type definitions for mode quantization in
volume $V$
\be
\hat{\vec A}(t,\vec x)
&=&
\sum_{s,\vec\kappa}
\sqrt{\frac{\hbar}{2\omega_{\vec \kappa} V}}
\Big(a_{s,\vec\kappa}
e^{-i\omega_{\vec \kappa} t} \vec e_{s,\vec\kappa}
e^{i\vec \kappa\cdot \vec x}
+
a^{\dag}_{s,\vec\kappa}e^{i\omega_{\vec \kappa} t} 
\vec e^{\,*}_{s,\vec\kappa}
e^{-i\vec \kappa\cdot \vec x}
\Big)\\
&=&
e^{iP\cdot x/\hbar} \hat{\vec A} e^{-iP\cdot x/\hbar}\\
\hat{\vec E}(t,\vec x)
&=&
i\sum_{s,\vec\kappa }
\sqrt{\frac{\hbar\omega_{\vec \kappa}}{2V}}
\Big(
a_{s,\vec\kappa}e^{-i\omega_{\vec \kappa} t} 
e^{i\vec \kappa\cdot \vec x}
\vec e_{s,\vec\kappa}
-
a^{\dag}_{s,\vec\kappa}e^{i\omega_{\vec \kappa} t} 
e^{-i\vec \kappa\cdot \vec x}
\vec e^{\,*}_{s,\vec\kappa}
\Big)\\
&=&
e^{iP\cdot x/\hbar} \hat{\vec E} e^{-iP\cdot x/\hbar}\\
\\
\hat{\vec B}(t,\vec x)
&=&
i\sum_{s,\vec\kappa }
\sqrt{\frac{\hbar\omega_{\vec \kappa}}{2V}}
\vec n_{\kappa}
\times 
\Big(a_{s,\vec\kappa}e^{-i\omega_{\vec \kappa} t} 
e^{i\vec \kappa\cdot \vec x}
\vec e_{s,\vec\kappa}
-
a^{\dag}_{s,\vec\kappa}e^{i\omega_{\vec \kappa} t} 
e^{-i\vec \kappa\cdot \vec x}
\vec e^{\,*}_{s,\vec\kappa}
\Big)\\
&=&
e^{iP\cdot x/\hbar} \hat{\vec B} e^{-iP\cdot x/\hbar},
\ee
where
\be
a_{s,\vec \kappa} &=&
|s,\vec \kappa\rangle \langle s,\vec \kappa|
\otimes a\label{a}\\
a^{\dag}_{s,\vec \kappa} &=&
|s,\vec \kappa\rangle \langle s,\vec \kappa|
\otimes a^{\dag}.\label{a^dag}
\ee
For later purposes we introduce the notation
\be
[a_{s,\vec \kappa},a_{s,\vec \kappa}^{\dag}]
=
1_{s,\vec \kappa}
=
|s,\vec \kappa\rangle\langle s,\vec \kappa|\otimes 
\bbox 1.\label{1_s}
\ee
Now take a state (say, in the Heisenberg picture)
\be
|\Psi\rangle
&=&
\sum_{s,\vec \kappa,n}\Psi_{s,\vec \kappa,n}
|s,\vec \kappa,n\rangle\\
&=&
\sum_{s,\vec \kappa}\Phi_{s,\vec \kappa}
|s,\vec \kappa\rangle|\alpha_{s,\vec \kappa}\rangle
\ee
where $|\alpha_{s,\vec \kappa}\rangle$ form a family of 
one-oscillator coherent states:
\be
a|\alpha_{s,\vec \kappa}\rangle
=
\alpha_{s,\vec \kappa}
|\alpha_{s,\vec \kappa}\rangle
\ee
The averages of the field operators are 
\be
\langle\Psi|\hat{\vec A}(t,\vec x)|\Psi\rangle
&=&
\sum_{s,\vec\kappa }|\Phi_{s,\vec \kappa}|^2
\sqrt{\frac{\hbar}{2\omega_{\vec \kappa} V}}
\Big(
\alpha_{s,\vec\kappa}
e^{-i\kappa\cdot x} 
\vec e_{s,\vec\kappa}
+
\alpha^*_{s,\vec\kappa}
e^{i\kappa\cdot x}
\vec e^{\,*}_{s,\vec\kappa}
\Big)\\
\langle\Psi|\hat{\vec E}(t,\vec x)|\Psi\rangle
&=&
\sum_{s,\vec\kappa }|\Phi_{s,\vec \kappa}|^2
\sqrt{\frac{\hbar\omega_{\vec \kappa}}{2V}}
\Big(
\alpha_{s,\vec\kappa}
e^{-i\kappa\cdot x}
\vec e_{s,\vec\kappa}
-
\alpha^*_{s,\vec\kappa}
e^{i\kappa\cdot x}
\vec e^{\,*}_{s,\vec\kappa}
\Big)\\
\langle\Psi|\hat{\vec B}(t,\vec x)|\Psi\rangle
&=&
i\sum_{s,\vec\kappa }|\Phi_{s,\vec \kappa}|^2
\sqrt{\frac{\hbar\omega_{\vec \kappa}}{2V}}
\Big(
\alpha_{s,\vec\kappa}e^{-i\kappa\cdot x}
\vec n_{\kappa}
\times 
\vec e_{s,\vec\kappa}
-
\alpha^*_{s,\vec\kappa}e^{i\kappa\cdot x}
\vec n_{\kappa}
\times 
\vec e^{\,*}_{s,\vec\kappa}
\Big)
\ee
These are just the classical fields. More precisely, the fields look
like averages 
of monochromatic coherent states with probabilities 
$|\Phi_{s,\vec \kappa}|^2$. The energy-momentum operators
satisfy also the standard relations 
\be
H
&=&
\frac{1}{2}
\int_V d^3x
\Big(
\hat{\vec E}(t,\vec x)\cdot \hat{\vec E}(t,\vec x)
+
\hat{\vec B}(t,\vec x)\cdot \hat{\vec B}(t,\vec x)
\Big),\label{P1}\\
\vec P
&=&
\int_V d^3x \hat{\vec E}(t,\vec x)\times \hat{\vec B}(t,\vec x).
\label{P2}
\ee
To end this section let us note that 
\be
\langle \Psi|H|\Psi\rangle
&=&
\sum_{s,\vec\kappa }
\hbar\omega_{\vec \kappa} 
|\Phi_{s,\vec\kappa}|^2
\Big(
|\alpha_{s,\vec\kappa}|^2
+\frac{1}{2}
\Big)\\
\langle \Psi|\vec P|\Psi\rangle
&=&
\sum_{s,\vec\kappa }
\hbar\vec \kappa
|\Phi_{s,\vec\kappa}|^2
\Big(
|\alpha_{s,\vec\kappa}|^2
+\frac{1}{2}
\Big).
\ee
The contribution from the vacuum fluctuations is nonzero but {\it finite\/}. 
One can phrase the latter property also as follows. The noncanonical algebra 
of creation-annihilation operators satisfies the resolution of identity 
\be
\sum_{s,\vec\kappa }
[a_{s,\vec \kappa},a_{s,\vec \kappa}^{\dag}]
=
\bbox 1\label{r-id}
\ee
wheras the canonical algebra would impliy 
\be
\sum_{s,\vec\kappa }
[a_{s,\vec \kappa},a_{s,\vec \kappa}^{\dag}]
=
\infty \bbox 1.
\ee

\section{``Second quantization''}

The Hilbert space of states of the field we have constructed is
spanned by vectors $|s,\vec \kappa,n\rangle$. Still there is no 
doubt that both in reality (and the standard formalism) 
there exist multiparticle entangled states such
as those spanned by tensor products of the form
\be
|+,\vec \kappa_1,1\rangle
|-,\vec \kappa_2,1\rangle,
\ee
and the similar. It seems that there is no reason to limit our
discussion to a {\it single\/} Hilbert space of a {\it single\/}
oscillator. What we have done so far was a quantization of the
electromagnetic field at the level of a ``one-particle" Hilbert
space.
Similarly to quantization of other physical systems 
the next step is to consider many particles. What is not obvious (physically)
is whether the oscillators should be considered as noninteracting.
This physical freedom leads to two natural candidates for a free-field 
Hamiltonian.  

The {\it noninteracting\/} extension is essentially clear. 
Having the one-particle
energy-momentum operators $P_a$  (i.e. generators of 4-translations in the 
1-particle Hilbert space) we define in the standard way their extensions 
to the Fock-type space
\be
{\cal P}_a
&=&P_a\nonumber\\
&\pp =&
\oplus\big(P_a\otimes\bbox 1+\bbox 1\otimes P_a\big)\nonumber\\
&\pp =&
\oplus \big(P_a\otimes\bbox 1\otimes\bbox 1
+\bbox 1\otimes P_a\otimes\bbox 1
+\bbox 1\otimes \bbox 1\otimes P_a\big)\nonumber\\
&\pp =&
\oplus\dots.
\ee
The $x$-dependence of fields is introduced similarly to the
one-particle level
\be
\vec {\cal F}(t,\vec x)
&=&
e^{i{\cal P}\cdot x/\hbar}
\vec {\cal F}
e^{-i{\cal P}\cdot x/\hbar}
\ee
but the field itself has yet to be defined. Assume 
\be
\vec {\cal F}
&=&
c_1\vec F\nonumber\\
&\pp =&
\oplus c_2\big(\vec F\otimes\bbox 1+\bbox 1\otimes \vec F\big)\nonumber\\
&\pp =&
\oplus c_3\big(\vec F\otimes\bbox 1\otimes\bbox 1
+\bbox 1\otimes \vec F\otimes\bbox 1
+\bbox 1\otimes \bbox 1\otimes \vec F\big)\nonumber\\
&\pp =&
\oplus\dots
\ee
where $c_k$ are constants discussed below, 
and $\vec F$ is $\hat {\vec A}$, $\hat {\vec E}$, or $\hat {\vec B}$.
The multi-oscillator annihilation operator associated with such fields 
must be therefore of the form 
\be
\bbox a_{s,\vec \kappa}
&=&
c_1a_{s,\vec \kappa}\nonumber\\
&\pp =&
\oplus c_2\big(a_{s,\vec \kappa}\otimes\bbox 1+
\bbox 1\otimes a_{s,\vec \kappa}\big)\nonumber\\
&\pp =&
\oplus c_3\big(a_{s,\vec \kappa}\otimes\bbox 1\otimes\bbox 1
+\bbox 1\otimes a_{s,\vec \kappa}\otimes\bbox 1
+\bbox 1\otimes \bbox 1\otimes a_{s,\vec \kappa}\big)\nonumber\\
&\pp =&
\oplus\dots.
\ee

Having two 1-particle operators, say $X$ and $Y$, one can easily 
establish a relation between the 1-particle commutator $[X,Y]$ and the 
commutator of the extensions $\cal X$, $\cal Y$:
\be
{[{\cal X},{\cal Y}]}
&=&c_1^2[X,Y]\nonumber\\
&\pp =&
\oplus c_2^2\big([X,Y]\otimes\bbox 1+\bbox 1\otimes [X,Y]\big)\nonumber\\
&\pp =&
\oplus c_3^2\big([X,Y]\otimes\bbox 1\otimes\bbox 1
+\bbox 1\otimes [X,Y]\otimes\bbox 1
+\bbox 1\otimes \bbox 1\otimes [X,Y]\big)\nonumber\\
&\pp =&
\oplus\dots.
\ee
The annihilation operators so defined satisfy therefore the algebra
\be
{[\bbox a_{s,\vec \kappa},\bbox a_{s',\vec \kappa\,'}^{\dag}]}
&=&0 \quad {\rm for}\,  (s,\vec \kappa)\neq (s',\vec \kappa\,'),\\
{[\bbox a_{s,\vec \kappa},\bbox a_{s,\vec \kappa\,}^{\dag}]}
&=&\bbox 1_{s,\vec \kappa},\\
{[\bbox a_{s,\vec \kappa},\bbox a_{s',\vec \kappa\,'}]}
&=&0\\
{[\bbox a_{s,\vec \kappa}^{\dag},\bbox a_{s',\vec \kappa\,'}^{\dag}]}
&=&0
\ee 
where the operator $\bbox 1_{s,\vec \kappa}$ is defined by
\be
\bbox 1_{s,\vec \kappa}
&=&c_1^2 1_{s,\vec \kappa}\nonumber\\
&\pp =&
\oplus c_2^2\big(1_{s,\vec \kappa}
\otimes\bbox 1+\bbox 1\otimes 1_{s,\vec \kappa}\big)\nonumber\\
&\pp =&
\oplus c_3^2\big(1_{s,\vec \kappa}
\otimes\bbox 1\otimes\bbox 1
+\bbox 1\otimes 1_{s,\vec \kappa}
\otimes\bbox 1
+\bbox 1\otimes \bbox 1\otimes 1_{s,\vec \kappa}\big)\nonumber\\
&\pp =&
\oplus\dots,
\ee
and $1_{s,\vec \kappa}$ is a single-oscillator operator (\ref{1_s}). 

An important property of the 
1-oscillator description was the resolution of identity (\ref{r-id}). The 
requirement that the same be valid at the multi oscillator level leads to 
$c_n=1/\sqrt{n}$.
In such a case one finds that 
\be
\bbox 1_{s,\vec \kappa}^2\neq \bbox 1_{s,\vec \kappa}
\ee
but nevertheless 
\be
\sum_{s,\vec \kappa}
\bbox 1_{s,\vec \kappa}=\bbox 1,\label{r of i}
\ee
that is $\bbox 1_{s,\vec \kappa}$ are the so-called positive operator valued 
(POV) measures \cite{Busch}. 
Below we shall give still another justification of this particular choice of 
$c_n$. 

We can finally write 
\be
\vec{\cal A}(t,\vec x)
&=&
\sum_{s,\vec\kappa}
\sqrt{\frac{\hbar}{2\omega_{\vec \kappa} V}}
\Big(\bbox a_{s,\vec\kappa}
e^{-i\omega_{\vec \kappa} t} \vec e_{s,\vec\kappa}
e^{i\vec \kappa\cdot \vec x}
+
\bbox a^{\dag}_{s,\vec\kappa}e^{i\omega_{\vec \kappa} t} 
\vec e^{\,*}_{s,\vec\kappa}
e^{-i\vec \kappa\cdot \vec x}
\Big)\\
&=&
e^{i{\cal P}\cdot x/\hbar} \vec{\cal A} e^{-i{\cal P}\cdot x/\hbar}\\
\vec{\cal E}(t,\vec x)
&=&
i\sum_{s,\vec\kappa }
\sqrt{\frac{\hbar\omega_{\vec \kappa}}{2V}}
\Big(
\bbox a_{s,\vec\kappa}e^{-i\omega_{\vec \kappa} t} 
e^{i\vec \kappa\cdot \vec x}
\vec e_{s,\vec\kappa}
-
\bbox a^{\dag}_{s,\vec\kappa}e^{i\omega_{\vec \kappa} t} 
e^{-i\vec \kappa\cdot \vec x}
\vec e^{\,*}_{s,\vec\kappa}
\Big)\\
&=&
e^{i{\cal P}\cdot x/\hbar} \vec{\cal E} e^{-i{\cal P}\cdot x/\hbar}\\
\\
\vec{\cal B}(t,\vec x)
&=&
i\sum_{s,\vec\kappa }
\sqrt{\frac{\hbar\omega_{\vec \kappa}}{2V}}
\vec n_{\kappa}
\times 
\Big(\bbox a_{s,\vec\kappa}e^{-i\omega_{\vec \kappa} t} 
e^{i\vec \kappa\cdot \vec x}
\vec e_{s,\vec\kappa}
-
\bbox a^{\dag}_{s,\vec\kappa}e^{i\omega_{\vec \kappa} t} 
e^{-i\vec \kappa\cdot \vec x}
\vec e^{\,*}_{s,\vec\kappa}
\Big)\\
&=&
e^{i{\cal P}\cdot x/\hbar} \vec{\cal B} e^{-i{\cal P}\cdot x/\hbar}.
\ee
These operators form a basis of the modified version of
nonrelativistic quantum optics.

Let us return for the moment to the case of a general $c_n$. 
A straightforward calculation shows that
\be
{\bf H}&=&
\frac{1}{2}
\int_V
d^3x
\Big(
\vec{\cal E}(t,\vec x)
\cdot
\vec{\cal E}(t,\vec x)
+
\vec{\cal B}(t,\vec x)
\cdot
\vec{\cal B}(t,\vec x)
\Big)
=
\frac{1}{2}\sum_{s,\vec\kappa}\hbar \omega_{\vec \kappa}
\Big(\bbox a_{s,\vec\kappa}^{\dag}\bbox a_{s,\vec\kappa} 
+\bbox a_{s,\vec\kappa} \bbox a_{s,\vec\kappa}^{\dag}\Big)\\
&=&
\sum_{s,\vec\kappa}\hbar \omega_{\vec \kappa}
\Bigg[
c_1^2\textstyle{\frac{1}{2}}
\{a_{s,\vec \kappa},a^{\dag}_{s,\vec \kappa}\}
\nonumber\\
&\pp =&
\pp{
\sum_{s,\vec\kappa}\hbar \omega_{\vec \kappa}
\Bigg[}
\oplus c_2^2\Big(\textstyle{\frac{1}{2}}
\{a_{s,\vec \kappa},a^{\dag}_{s,\vec \kappa}\}
\otimes\bbox 1+
\bbox 1\otimes \textstyle{\frac{1}{2}}
\{a_{s,\vec \kappa},a^{\dag}_{s,\vec \kappa}\}
+
a^{\dag}_{s,\vec \kappa}\otimes a_{s,\vec \kappa}
+
a_{s,\vec \kappa}\otimes a^{\dag}_{s,\vec \kappa}
\Big)\nonumber\\
&\pp =&
\pp{
\sum_{s,\vec\kappa}\hbar \omega_{\vec \kappa}
\Bigg[}
\oplus c_3^2
\Big(\textstyle{\frac{1}{2}}
\{a_{s,\vec \kappa},a^{\dag}_{s,\vec \kappa}\}\otimes\bbox 1\otimes\bbox 1
+
\bbox 1\otimes\textstyle{\frac{1}{2}} 
\{a_{s,\vec \kappa},a^{\dag}_{s,\vec \kappa}\}\otimes\bbox 1
+
\bbox 1\otimes \bbox 1\otimes\textstyle{\frac{1}{2}}
 \{a_{s,\vec \kappa},a^{\dag}_{s,\vec \kappa}\}
\nonumber\\
&\pp =&\pp +
\pp{
\sum_{s,\vec\kappa}\hbar \omega_{\vec \kappa}
\Bigg[c_3(}
+
a_{s,\vec \kappa}\otimes a^{\dag}_{s,\vec \kappa}\otimes\bbox 1
+
 a^{\dag}_{s,\vec \kappa}\otimes a_{s,\vec \kappa}\otimes\bbox 1
+
\bbox 1\otimes a_{s,\vec \kappa}\otimes  a^{\dag}_{s,\vec \kappa}\nonumber\\
&\pp =&\pp +
\pp{
\sum_{s,\vec\kappa}\hbar \omega_{\vec \kappa}
\Bigg[c_3(}
+
\bbox 1\otimes  a^{\dag}_{s,\vec \kappa}\otimes a_{s,\vec \kappa}
+
 a^{\dag}_{s,\vec \kappa}\otimes \bbox 1\otimes a_{s,\vec \kappa}
+a_{s,\vec \kappa}\otimes\bbox 1\otimes  a^{\dag}_{s,\vec \kappa}
\Big)\nonumber\\
&\pp =&
\pp{
\sum_{s,\vec\kappa}\hbar \omega_{\vec \kappa}
\Bigg[}
\oplus\dots
\Bigg]
\ee
where $\{\cdot,\cdot\}$ denotes the anti-commutator.
Comparing this with the generator of time translations 
\be
{\cal H}=c{\cal P}_0
&=&
\sum_{s,\vec\kappa}\hbar \omega_{\vec \kappa}
\Bigg[
\textstyle{\frac{1}{2}}
\{a_{s,\vec \kappa},a^{\dag}_{s,\vec \kappa}\}
\nonumber\\
&\pp =&
\pp{
\sum_{s,\vec\kappa}\hbar \omega_{\vec \kappa}
\Bigg[}
\oplus \Big(
\textstyle{\frac{1}{2}}
\{a_{s,\vec \kappa},a^{\dag}_{s,\vec \kappa}\}
\otimes\bbox 1+
\bbox 1\otimes\textstyle{\frac{1}{2}} 
\{a_{s,\vec \kappa},a^{\dag}_{s,\vec \kappa}\}
\Big)\nonumber\\
&\pp =&
\pp{
\sum_{s,\vec\kappa}\hbar \omega_{\vec \kappa}
\Bigg[}
\oplus 
\Big(
\textstyle{\frac{1}{2}}
\{a_{s,\vec \kappa},a^{\dag}_{s,\vec \kappa}\}\otimes\bbox 1\otimes\bbox 1
+
\bbox 1\otimes\textstyle{\frac{1}{2}}
 \{a_{s,\vec \kappa},a^{\dag}_{s,\vec \kappa}\}\otimes\bbox 1
+
\bbox 1\otimes \bbox 1\otimes 
\textstyle{\frac{1}{2}}
\{a_{s,\vec \kappa},a^{\dag}_{s,\vec \kappa}\}
\Big)\nonumber\\
&\pp =&
\pp{
\sum_{s,\vec\kappa}\hbar \omega_{\vec \kappa}
\Bigg[}
\oplus\dots
\Bigg]
\ee
we can see that there {\it is\/} 
a relation between $\cal H$ and $\bf H$ but the latter
contains terms describing interactions between the oscillators. 
The contribution from these interactions vanishes on vacuum states. 
Below, when we introduce the notion of a generalized coherent state, we will
be able to relate averages of  $\cal H$ and $\bf H$. 
In a similar way one can introduce the ``Pointing operator''
\be
\vec{\bf P} &=&
\int_V
d^3x\,
\vec{\cal E}(t,\vec x)
\times
\vec{\cal B}(t,\vec x)
=
\frac{1}{2}\sum_{s,\vec\kappa}\hbar \vec\kappa
\Big(\bbox a_{s,\vec\kappa}^{\dag}\bbox a_{s,\vec\kappa} 
+\bbox a_{s,\vec\kappa} \bbox a_{s,\vec\kappa}^{\dag}\Big).
\ee
Its relation to the generator of 3-translations $\vec {\cal P}$ is similar 
to this between $\cal H$ and $\bf H$. 

In the above construction the only element which is beyond a simple 
transition to many oscillators is the choice of $c_n$. For different 
choices of these constants we obtain different 
representations of non-CCR
and therefore also different quantization schemes. 
Several different ways of reasoning lead to $c_n=1/\sqrt{n}$ as we shall 
also see in the next sections. 

\section{Some particular states}

We assume that all the multi-oscillator states are symmetric with
respect to permutations of the oscillators. 

\subsection{Generalized coherent states}

For general $c_n$ an eigenstate of $\bbox a_{s,\vec\kappa}$ 
corresponding to the eigenvalue $\alpha_{s,\vec \kappa}$
is of the form
\be
|\bbox \alpha_{s,\vec \kappa}\rangle
&=&
f_1(s,\vec \kappa)|s,\vec \kappa,\alpha_{s,\vec \kappa}/c_1\rangle\nonumber\\
&\pp =&
\oplus
f_2(s,\vec \kappa)
|s,\vec \kappa,\alpha_{s,\vec \kappa}/(2c_2)\rangle
|s,\vec \kappa,\alpha_{s,\vec \kappa}/(2c_2)\rangle\nonumber\\
&\pp =&
\oplus
f_3(s,\vec \kappa)
|s,\vec \kappa,\alpha_{s,\vec \kappa}/(3c_3)\rangle
|s,\vec \kappa,\alpha_{s,\vec \kappa}/(3c_3)\rangle
|s,\vec \kappa,\alpha_{s,\vec \kappa}/(3c_3)\rangle\nonumber\\
&\pp =&
\oplus\dots
\ee
where 
\be
|s,\vec \kappa,\alpha_{s,\vec \kappa}\rangle=|s,\vec \kappa\rangle
|\alpha_{s,\vec \kappa}\rangle,
\ee
$\sum_{n=1}^\infty|f_n(s,\vec \kappa)|^2=1$,
and 
$a |\alpha_{s,\vec \kappa}\rangle=\alpha_{s,\vec \kappa}
|\alpha_{s,\vec \kappa}\rangle$. 
What is interesting not all $f_n$ have to be nonvanishing. 

The average ``energies'' of the field in the above eigenstate are
\be
\langle\bbox \alpha_{s,\vec \kappa}|
{\cal H}
|\bbox \alpha_{s,\vec \kappa}\rangle
&=&
\hbar\omega_{\vec \kappa}
|\alpha_{s,\vec \kappa}|^2
\sum_{n=1}^\infty\frac{1}{nc_n^2}|f_n(s,\vec \kappa)|^2
+
\frac{1}{2}\hbar\omega_{\vec \kappa}\sum_{n=1}^\infty n|f_n(s,\vec \kappa)|^2
\ee
and 
\be
\langle\bbox \alpha_{s,\vec \kappa}|
{\bf H}
|\bbox \alpha_{s,\vec \kappa}\rangle
&=&
\hbar\omega_{\vec \kappa}
|\alpha_{s,\vec \kappa}|^2
+
\frac{1}{2}\hbar\omega_{\vec \kappa}
\sum_{n=1}^\infty nc_n^2|f_n(s,\vec \kappa)|^2.
\ee
The two averages will differ only by the value of the vacuum contribution 
if $c_n=1/\sqrt{n}$ which leads us back to the above mentioned 
choice of $c_n$. 
With this choice and taking the general combination of coherent states 
\be
|\bbox \Psi\rangle
=\sum_{s,\vec \kappa}\Phi_{s,\vec \kappa}|\bbox \alpha_{s,\vec \kappa}\rangle
\ee
we find 
\be
\langle\bbox \Psi|
{\cal H}
|\bbox \Psi\rangle
&=&
\sum_{s,\vec \kappa}
\hbar\omega_{\vec \kappa}
|\Phi_{s,\vec \kappa}|^2
|\alpha_{s,\vec \kappa}|^2
+
\frac{1}{2}
\sum_{s,\vec \kappa}
\hbar\omega_{\vec \kappa}
|\Phi_{s,\vec \kappa}|^2
\sum_{n=1}^\infty n|f_k(s,\vec \kappa)|^2
\ee
and 
\be
\langle \bbox \Psi|
{\bf H}
|\bbox \Psi\rangle
&=&
\sum_{s,\vec \kappa}\hbar\omega_{\vec \kappa}
|\Phi_{s,\vec \kappa}|^2|\alpha_{s,\vec \kappa}|^2
+
\frac{1}{2}\sum_{s,\vec \kappa}
\hbar\omega_{\vec \kappa}|\Phi_{s,\vec \kappa}|^2.
\ee

One may wonder, then, what is the more natural choice of the free-field 
Hamiltonian: ${\bf H}$ describing interacting oscillators, 
or $\cal H$ describing the noninteracting ones? The coherent-state average of 
${\bf H}$ does not depend on the average number of the oscillators 
and naturally includes the process of energy exchange between different 
oscillators. Moreover, both ${\bf H}$ and $\vec{\bf P}$ are defined 
in the standard way in terms of the multi-oscillator non-CCR algebra.
With this choice of free dynamics we find 
\be
e^{i{\bf H}t/\hbar}\bbox a_{s,\vec \kappa}e^{-i{\bf H}t/\hbar}
=
e^{-i\omega_{\vec \kappa} t\bbox 1_{s,\vec \kappa}} \bbox a_{s,\vec \kappa}
\ee
as opposed to the standard formula
\be
e^{i{\cal H}t/\hbar}\bbox a_{s,\vec \kappa}e^{-i{\cal H}t/\hbar}
=
e^{-i\omega_{\vec \kappa} t} \bbox a_{s,\vec \kappa}.
\ee
The latter choice is simpler because it leads to the standard form of the 
interaction-picture Hamiltonian and therefore will be the basis of our
non-canonical quantum optics. The version based on  ${\bf H}$ is a subject
of ongoing study. 
\subsection{Vacuum}

Similarly to the one-oscillator case the traditional notion of a
vacuum state is replaced in our formalism by a vacuum {\it
subspace\/} consisting of all the vectors annihilated by all 
$\bbox a_{s,\vec\kappa}$. Their general form is 
\be
|\bbox 0\rangle
&=&
\sum_{s,\vec \kappa}O^{(1)}_{s,\vec \kappa,0}
|s,\vec \kappa,0\rangle\nonumber\\
&\pp =&
\oplus
\sum_{s_j,\vec \kappa_{j}}O^{(2)}_{s_1,s_2,\vec
\kappa_{1},\vec\kappa_{2},0,0}
|s_1,\vec \kappa_{1},0\rangle
|s_2,\vec \kappa_{2},0\rangle\nonumber\\
&\pp =&
\oplus
\sum_{s_j,\vec \kappa_{\lambda_j}}O^{(3)}_{s_1,s_2,s_3\vec
\kappa_{\lambda_1},\vec\kappa_{\lambda_2},\vec\kappa_{\lambda_3},0,0,0}
|s_1,\vec \kappa_{1},0\rangle
|s_2,\vec \kappa_{2},0\rangle
|s_3,\vec \kappa_{3},0\rangle\nonumber\\
&\pp =&
\oplus\dots
\ee
It seems that there is no reason for introducing the standard 
unique ``vacuum state"
understood as the cyclic vector of the GNS construction. 

As an example of a vacuum state consider
\be
|\bbox 0\rangle
&=&
\sqrt{p_1}|O\rangle\nonumber\\
&\pp =&
\oplus
\sqrt{p_2}|O\rangle|O\rangle
\nonumber\\
&\pp =&
\oplus
\sqrt{p_3}
|O\rangle|O\rangle|O\rangle\nonumber\\
&\pp =&
\oplus\dots\label{multi-vac}
\ee
The average energy of the free-field vacuum state is therefore
\be
\overline{{\cal H}}=
\langle \bbox 0|
{\cal H}
|\bbox 0\rangle
=
\sum_{n=1}^\infty
np_n \langle O|
H
|O\rangle=\overline{n} \overline{H} 
\ee
where $\overline{n}$ and $\overline{H}$ are, respectively, the
average number of oscillators and the average energy of a single
oscillator. For the sake of completeness let us note that 
\be
\overline{{\bf H}}=
\langle \bbox 0|
{\bf H}
|\bbox 0\rangle
=
\sum_{n=1}^\infty
c_n^2np_n \langle O|
H
|O\rangle=\overline{H}\sum_{n=1}^\infty
c_n^2np_n 
\ee
For $c_n=1$ we obtain $\overline{{\bf H}}=\overline{{\cal H}}$; 
for $c_n=1/\sqrt{n}$ $\overline{{\bf H}}=\overline{H}$ the latter being 
independent of the number of oscillators.
In both cases no problem with infinite vacuum energy is found. 
Obviously, one can contemplate also other vacua, say, in
entangled or mixed states.

\subsection{Multi-photon states}

Assume $|\bbox 0\rangle$ is a vacuum state.
The non-canonical algebra (non-CCR) differes from the canonical one in the
commutator
\be
[\bbox a_\lambda,\bbox a_\lambda^{\dag}]=\bbox 1_\lambda
\ee
where for any $\lambda$, $\lambda'$
\be
[\bbox 1_{\lambda'},\bbox a_\lambda]=[\bbox 1_{\lambda'},\bbox
a_\lambda^{\dag}]=0
\ee
where $\lambda$ stands for $(s,\vec\kappa)$. 
A normalized state describing a collection of photons is defined in analogy 
to the standard formalism by
\be
\frac{1}{\sqrt{
n_1!n_2!\dots n_N!
\langle \bbox 0|
\bbox 1_{\lambda_N}^{n_N}
\dots
\bbox 1_{\lambda_2}^{n_2}
\bbox 1_{\lambda_1}^{n_1}
|\bbox 0\rangle}}
(\bbox a^{\dag}_{\lambda_N})^{n_N}
\dots
(\bbox a^{\dag}_{\lambda_2})^{n_2}
(\bbox a^{\dag}_{\lambda_1})^{n_1}
|\bbox 0\rangle
=:
|\bbox n_{\lambda_1},\bbox n_{\lambda_2},\dots, \bbox n_{\lambda_N}\rangle
\ee
States corresponding to the same $\lambda$ but different $n$'s, or to the
same $n$ but different $\lambda$'s, are orthogonal. 
As a consequence a non-canonical vacuum average of any product of 
non-canonical creation and
annihilation operators vanishes if and only if an analogous expression 
formulated in terms of the canonical objects does. 
This property is a consequence of three facts which hold true in both
formalisms: (a) annihilation operators annihilate vacuum states, (b) 
creation operators are obtained by a Hermitian conjugate of the annihilation 
operators, and (c) the RHS of a commutator of creation and annihilation
operators commutes with all creation and annihilation operators. 

\section{Perturbation theory}

It is essential that, similarly to the one-oscillator formalism, the free
Hamiltonian (defined simply as a generator of time translations) 
generates the standard  dynamics of annihilation operators:
\be
e^{i{\cal H}t/\hbar}\bbox a_{s,\vec \kappa}e^{-i{\cal H}t/\hbar}
=
e^{-i\omega_{\vec \kappa} t} \bbox a_{s,\vec \kappa}.
\ee
Accordingly, the form of the interaction-picture Hamiltonian will by the
same as in the standard theory. This would not be quite the same if we have 
chosen $\bf H$ in the role of the free Hamiltonian (an option which, 
nevertheless, should be investigated).
In what follows we start with 
$H=H_0+V$, where 
\be
H_0 &=& 
H_A
+
{\cal H}\\
V &=&
-\frac{e}{m}
\vec{\cal A}(\vec x)\cdot \vec p
\nonumber\\
&=&
-\frac{e}{m}
\sum_{s,\vec\kappa}
\sqrt{\frac{\hbar}{2\omega_{\vec\kappa} V}}
\Big(\bbox a_{s,\vec\kappa}
e^{i\vec \kappa\cdot \vec x}
\vec e_{s,\vec\kappa}\cdot\vec p
+
\bbox a^{\dag}_{s,\vec\kappa}
e^{-i\vec \kappa\cdot \vec x}
\vec e^{\,*}_{s,\vec\kappa}\cdot\vec p
\Big).
\ee
In the interaction picture we get 
\be
V(t) &=&
-\frac{e}{m}
\vec{\cal A}(t,\vec x)\cdot \vec p(t)
\nonumber\\
&=&
-\frac{e}{m}
\sum_{s,\vec\kappa}
\sqrt{\frac{\hbar}{2\omega_{\vec\kappa} V}}
\Big(\bbox a_{s,\vec\kappa}
e^{-i(\omega_{\vec\kappa}t-\vec \kappa\cdot \vec x)}
\vec e_{s,\vec\kappa}\cdot\vec p(t)
+
\bbox a^{\dag}_{s,\vec\kappa}
e^{i(\omega_{\vec\kappa}t-\vec \kappa\cdot \vec x)}
\vec e^{\,*}_{s,\vec\kappa}\cdot\vec p(t)
\Big),
\ee
and 
\be
\vec p(t)=e^{iH_At}\vec p\,e^{-iH_At}.
\ee
Since we are purposefully neglecting the ``$\vec A\,^2$" term in the
Hamiltonian, one should restrict the analysis to the dipole
approximation and therefore it is justified to set $\vec x=0$:
\be
V(t) &=&
\sum_{\lambda}
\Big(\bbox a_{\lambda}
\hat g_{\lambda}(t)
+
\bbox a^{\dag}_{\lambda}
\hat g^{\dag}_{\lambda}(t)
\Big).
\ee
The operators 
\be
\hat g_{s,\vec\kappa}(t)=
-\frac{e}{m}\sqrt{\frac{\hbar}{2\omega_{\vec\kappa} V}}
e^{-i\omega_{\vec\kappa}t}\vec e_{s,\vec\kappa}\cdot\vec
p(t)=
\hat g_{\lambda}(t)
\ee
are identical to those from the standard formalism and act only on
atomic degrees of freedom (i.e. commute with $\bbox
a_{s,\vec\kappa}$).

\subsection{Spontaneous decay of an excited state}

The first problem we shall treat in the non-canonical way is a lifetime of
an excited atomic state. The problem, as we shall see, is of particular
importance for the physical interpretation of the non-canonical formalism. 

Assume that at $t=0$ the atom-field system is described by the state 
$|\Psi(0)\rangle=|\bbox 0,A\rangle$. The amplitude that the atom
remains in the excited state is 
\be
\langle \bbox 0, A|\Psi(t)\rangle
&=&
1
\nonumber\\
&\pp = &
%2222222222222222222222222222222222222222222222222222
+
\frac{1}{(i\hbar)^2}\int_{0}^{t}
dt_2
\int_{0}^{t_2}
dt_1\,
\sum_{\lambda_1\lambda_2}
\langle \bbox 0|
\bbox a_{\lambda_2}
\bbox a^{\dag}_{\lambda_1}
|\bbox 0\rangle
\langle A|
\hat g_{\lambda_2}(t_2)
\hat g^{\dag}_{\lambda_1}(t_1)
|A\rangle
\nonumber\\
&\pp = &
%4444444444444444444444444444444444444444444444444444
+
\frac{1}{(i\hbar)^4}\int_{0}^{t}
dt_4
\int_{0}^{t_4}
dt_3
\int_{0}^{t_3}
dt_2
\int_{0}^{t_2}
dt_1\,
\sum_{\lambda_1\lambda_2\lambda_3\lambda_4}
\langle \bbox 0|
\bbox a_{\lambda_4}
\bbox a^{\dag}_{\lambda_3}
\bbox a_{\lambda_2}
\bbox a^{\dag}_{\lambda_1}
|\bbox 0\rangle
\langle A|
\hat g_{\lambda_4}(t_4)
\hat g^{\dag}_{\lambda_3}(t_3)
\hat g_{\lambda_2}(t_2)
\hat g^{\dag}_{\lambda_1}(t_1)
|A\rangle
\nonumber\\
&\pp = &
+
\frac{1}{(i\hbar)^4}\int_{0}^{t}
dt_4
\int_{0}^{t_4}
dt_3
\int_{0}^{t_3}
dt_2
\int_{0}^{t_2}
dt_1\,
\sum_{\lambda_1\lambda_2\lambda_3\lambda_4}
\langle \bbox 0|
\bbox a_{\lambda_4}
\bbox a_{\lambda_3}
\bbox a^{\dag}_{\lambda_2}
\bbox a^{\dag}_{\lambda_1}
|\bbox 0\rangle
\langle A|
\hat g_{\lambda_4}(t_4)
\hat g_{\lambda_3}(t_3)
\hat g^{\dag}_{\lambda_2}(t_2)
\hat g^{\dag}_{\lambda_1}(t_1)
|A\rangle
\nonumber\\
&\pp = &
+\dots
\nonumber\\
&=&
1
\nonumber\\
&\pp = &
%2222222222222222222222222222222222222222222222222222
+
\frac{1}{(i\hbar)^2}\int_{0}^{t}
dt_2
\int_{0}^{t_2}
dt_1\,
\sum_{\lambda_1\lambda_2}
\langle \hat 0|
\hat a_{\lambda_2}
\hat a^{\dag}_{\lambda_1}
|\hat 0\rangle
\langle A|
\hat g_{\lambda_2}(t_2)
\hat g^{\dag}_{\lambda_1}(t_1)
|A\rangle X^{01}_{\lambda_2\lambda_1}
\nonumber\\
&\pp = &
%4444444444444444444444444444444444444444444444444444 
+
\frac{1}{(i\hbar)^4}\int_{0}^{t}
dt_4
\int_{0}^{t_4}
dt_3
\int_{0}^{t_3}
dt_2
\int_{0}^{t_2}
dt_1\,
\sum_{\lambda_1\lambda_2\lambda_3\lambda_4}
\langle \hat 0|
\hat a_{\lambda_4}
\hat a^{\dag}_{\lambda_3}
\hat a_{\lambda_2}
\hat a^{\dag}_{\lambda_1}
|\hat 0\rangle
\langle A|
\hat g_{\lambda_4}(t_4)
\hat g^{\dag}_{\lambda_3}(t_3)
\hat g_{\lambda_2}(t_2)
\hat g^{\dag}_{\lambda_1}(t_1)
|A\rangle 
\nonumber\\
&\pp =&\pp =\times
X^{0101}_{\lambda_4\lambda_3\lambda_2\lambda_1}
\nonumber\\
&\pp = &
+
\frac{1}{(i\hbar)^4}\int_{0}^{t}
dt_4
\int_{0}^{t_4}
dt_3
\int_{0}^{t_3}
dt_2
\int_{0}^{t_2}
dt_1\,
\sum_{\lambda_1\lambda_2\lambda_3\lambda_4}
\langle \hat 0|
\hat a_{\lambda_4}
\hat a_{\lambda_3}
\hat a^{\dag}_{\lambda_2}
\hat a^{\dag}_{\lambda_1}
|\hat 0\rangle
\langle A|
\hat g_{\lambda_4}(t_4)
\hat g_{\lambda_3}(t_3)
\hat g^{\dag}_{\lambda_2}(t_2)
\hat g^{\dag}_{\lambda_1}(t_1)
|A\rangle 
\nonumber\\
&\pp =&\pp =\times
X^{0011}_{\lambda_4\lambda_3\lambda_2\lambda_1}
\nonumber\\
&\pp = &
+\dots
\ee
In the above perturbative expansion we have explicitly shown all the
nonvanishing terms up to the fifth order of perturbation theory.
Here $\langle \hat 0|
\hat a_{\lambda_4}
\hat a_{\lambda_3}
\hat a^{\dag}_{\lambda_2}
\hat a^{\dag}_{\lambda_1}
|\hat 0\rangle$ etc. are the expressions one would have obtained in ordinary
canonical formalism and
\be
X^{0011}_{\lambda_4\lambda_3\lambda_2\lambda_1}
=
\frac{
\langle \bbox 0|
\bbox a_{\lambda_4}
\bbox a_{\lambda_3}
\bbox a^{\dag}_{\lambda_2}
\bbox a^{\dag}_{\lambda_1}
|\bbox 0\rangle
}{
\langle \hat 0|
\hat a_{\lambda_4}
\hat a_{\lambda_3}
\hat a^{\dag}_{\lambda_2}
\hat a^{\dag}_{\lambda_1}
|\hat 0\rangle
}
\ee
etc. Such expressions are well defined since whenever their denominators
vanish the whole term it containing vanishes as well. This 
fact is of crucial importance and shows that the
perturbative expansions in both canonical and noncanonical frameworks
contain terms of {\it exactly the same type\/} but differing by the numerical
factors $X_{\dots}^{\dots}$. 

Let us note that in the above calculation we have not used the
explicit realization of the non-canonical algebra but only the algebra
itself. At such a general level both the canonincal and non-canonical 
theories can be regarded as particular cases of a more general theory
characterized by the algebra 
\be
{[a_{\lambda},a^{\dag}_{\lambda'}]}&=&\delta_{\lambda\lambda'}I_\lambda,\\
{[a_{\lambda},a_{\lambda'}]}&=& 0,\\
{[a^{\dag}_{\lambda},a^{\dag}_{\lambda'}]}&=& 0,\\
{[a^{\dag}_{\lambda},I_{\lambda'}]} &=&0,\\
{[a_{\lambda},I_{\lambda'}]}&=&0.
\ee
The canonical choice, based on oscillators with classical parameter
$\lambda$,  is $I_{\lambda}=\bbox 1$; the choice 
based on oscillators with eigenvalue $\lambda$ is 
$I_{\lambda}=\bbox 1_{\lambda}$. 

To proceed further and get more information as to the physical meaning of 
the non-canonical dynamics we have to make the analysis less general.
First of all let us stick to the particular choice of 
$\bbox 1_{\lambda}$ in terms of POV measures we have introduced earlier 
and assume that $\sum_\lambda \bbox 1_{\lambda}=\bbox 1$. 
Second, let us take the vacuum state in the form (\ref{multi-vac}). 
Under such assumptions we can explicitly compute the factors 
$X_{\dots}^{\dots}$:
\be
X^{01}_{\lambda\lambda}
&=&
\frac{
\langle \bbox 0|
\bbox a_{\lambda}
\bbox a^{\dag}_{\lambda}
|\bbox 0\rangle
}{
\langle \hat 0|
\hat a_{\lambda}
\hat a^{\dag}_{\lambda}
|\hat 0\rangle
}
=
\langle \bbox 0|
\bbox 1_{\lambda}
|\bbox 0\rangle
=|O_\lambda|^2
\\
X^{0101}_{\lambda\lambda\lambda'\lambda'}
&=&
\frac{
\langle \bbox 0|
\bbox a_{\lambda}
\bbox a^{\dag}_{\lambda}
\bbox a_{\lambda'}
\bbox a^{\dag}_{\lambda'}
|\bbox 0\rangle
}{
\langle \hat 0|
\hat a_{\lambda}
\hat a^{\dag}_{\lambda}
\hat a_{\lambda'}
\hat a^{\dag}_{\lambda'}
|\hat 0\rangle
}
=
\langle \bbox 0|
\bbox 1_{\lambda}
\bbox 1_{\lambda'}
|\bbox 0\rangle
=
\sum_{n=1}^\infty p_n\Big(1- \frac{1}{n}\Big)|O_\lambda|^2|O_{\lambda'}|^2
=
\Big(1- \big\langle 1/n\big\rangle\Big)|O_\lambda|^2|O_{\lambda'}|^2
\\
X^{0101}_{\lambda\lambda\lambda\lambda}
&=&
\frac{
\langle \bbox 0|
\bbox a_{\lambda}
\bbox a^{\dag}_{\lambda}
\bbox a_{\lambda}
\bbox a^{\dag}_{\lambda}
|\bbox 0\rangle
}{
\langle \hat 0|
\hat a_{\lambda}
\hat a^{\dag}_{\lambda}
\hat a_{\lambda}
\hat a^{\dag}_{\lambda}
|\hat 0\rangle
}
=
\langle \bbox 0|
\bbox 1_{\lambda}^2
|\bbox 0\rangle
=
\Big(1- \big\langle 1/n\big\rangle\Big)|O_\lambda|^4
+
\big\langle 1/n\big\rangle|O_\lambda|^2
\\
X^{0011}_{\lambda\lambda\lambda\lambda}
&=&
\frac{
\langle \bbox 0|
\bbox a_{\lambda}
\bbox a_{\lambda}
\bbox a^{\dag}_{\lambda}
\bbox a^{\dag}_{\lambda}
|\bbox 0\rangle
}{
\langle \hat 0|
\hat a_{\lambda}
\hat a_{\lambda}
\hat a^{\dag}_{\lambda}
\hat a^{\dag}_{\lambda}
|\hat 0\rangle
}
=
\langle \bbox 0|
\bbox 1_{\lambda}^2
|\bbox 0\rangle
=
\Big(1- \big\langle 1/n\big\rangle\Big)|O_\lambda|^4
+
\big\langle 1/n\big\rangle|O_\lambda|^2
\\
X^{0011}_{\lambda\lambda'\lambda\lambda'}
&=&
\frac{
\langle \bbox 0|
\bbox a_{\lambda}
\bbox a_{\lambda'}
\bbox a^{\dag}_{\lambda}
\bbox a^{\dag}_{\lambda'}
|\bbox 0\rangle
}{
\langle \hat 0|
\hat a_{\lambda}
\hat a_{\lambda'}
\hat a^{\dag}_{\lambda}
\hat a^{\dag}_{\lambda'}
|\hat 0\rangle
}
=
\langle \bbox 0|
\bbox 1_{\lambda}
\bbox 1_{\lambda'}
|\bbox 0\rangle
=
\Big(1- \big\langle 1/n\big\rangle\Big)|O_\lambda|^2|O_{\lambda'}|^2.
\ee
What is interesting (and very characteristic) all these factors are smaller
than 1 (this follows trivially from $\sum_\lambda |O_\lambda|^2=1$). 
An analysis of higher order terms shows that this is a generic property of
the non-canonical perturbation theory. The $n$ occuring in the average 
$\langle 1/n\rangle$ is the number-of-oscillators operator. For realistic
vacua one may expect the average number of oscillators to be large
and therefore $\langle 1/n\rangle\approx 0$. Taking a more general vacuum
state we arrive at $\langle 1/n_\lambda\rangle$ instead of 
$\langle 1/n\rangle$, which means that the influence of the vacuum  
may vary from frequency to frequency (i.e. from point to point in space).

Now, if we ignore the corrections coming from $\langle 1/n\rangle$
we can see that the non-canonical perturbative expansion of the amplitude is
the same we would have obtained by using the {\it standard theory\/} but
with $\hat g_\lambda$ regularized by $\hat g_\lambda \to
O_\lambda \hat g_\lambda$. As a consequence there exists a
natural cut-off in the theory which follows {\it only\/} from the fact that the
vacuum wave function is square-integrable and therefore 
$O_{s,\vec\kappa}\to 0$ for 
$|\vec\kappa|\to \infty$. 
It is quite remarkable that the same 
mechanism that eliminated the infinite vacuum
energy plays a similar role in the other parts of the theory. 
As we shall see shortly the actual role of the
vacuum can be analyzed only in a fully relativistic setting since then the
charge and mass renormalization come into play. 

However, for the sake of concreteness and to make some rough estimates of the 
effects involved let us take the trivial example where 
the vacuum amplitudes are constant, say, $O_{s,\vec\kappa}=C$ for all 
$\omega_{\vec \kappa}< \omega_{\rm max}$ and zero otherwise.  
The dynamics of the amplitude is then (up to $\langle 1/n\rangle\approx 0$) 
equivalent to the standard one with the cut-off at $\omega_{\rm max}$ 
and the coupling constant $e/m$ replaced by $Ce/m$. This implies that $e$
and $m$ have to be treated as {\it bare\/} parameters and
the experimantal value is $e_{\rm ex}/m_{\rm ex}=Ce/m$. 
With this observation in mind we can discuss non-canonically 
other perturbative effects which are widely believed to be a consequence 
of the standard canonical quantization. Below we find it useful to make the bare 
parameter $e/m$ explicit in calculations and for this reason will use the notation 
$\hat g_\lambda=(e/m)\hat f_\lambda$.  

\subsection{Spontaneous emission of $N$ identical photons in $N$-th order perturbation
theory}

Assume the atomic transition is $|A\rangle\to |B\rangle$. 
Up to the $N$-th order perturbative correction we get
\be
\langle \bbox N_\lambda, B|\Psi(t)\rangle
&=&
\Big(\frac{e}{m}\Big)^N
\frac{1}{(i\hbar)^N}
\int_0^t dt_N\int_0^{t_N}\dots \int_0^{t_2}
dt_1
\langle \bbox N_\lambda|
\bbox a_\lambda^{\dag N}
|\bbox 0\rangle\langle B|
\hat f^{\dag}_{\lambda}(t_N)
\dots
\hat f^{\dag}_{\lambda}(t_1)
|A\rangle\nonumber\\
&=&
\Big(\frac{e}{m}\Big)^N
\frac{\langle \bbox N_\lambda|\bbox a_\lambda^{\dag N}|\bbox 0\rangle}
{\langle \hat N_\lambda|\hat a_\lambda^{\dag N}|\hat 0\rangle}
\times ({\rm relevant\,canonical\,formula})\nonumber\\
&=&
\Big(\frac{e}{m}\Big)^N
\sqrt{\langle \bbox 0|\bbox 1_\lambda^N|\bbox 0\rangle}
\times ({\rm relevant\,canonical\,formula})\nonumber\\
\ee
where the ``hatted'' expressions are those from the canonical theory.
As we can see the task is reduced to computing 
$\langle \bbox 0|\bbox 1_\lambda^N|\bbox 0\rangle$. The general formula, 
valid for any $N$, is somewhat complicated and not very illuminating. 
The cases $N=1$ and $N=2$ we have already met.  Making the simplifying choice 
of a very ``flat'' distribution of the vacuum modes 
(i.e.  $O_{s,\vec\kappa}=C$ below some threshold)
we get
\be
\langle \bbox 1_\lambda, B|\Psi(t)\rangle
&=&
C\langle \hat 1_\lambda, B|\hat \Psi(t)\rangle\\
\langle \bbox 2_\lambda, B|\Psi(t)\rangle
&=&
\sqrt{C^4 +\big\langle 1/n\big\rangle (C^2-C^4)}
\langle \hat 2_\lambda, B|\hat \Psi(t)\rangle.
\ee
Identifying  $Ce/m$ with $e_{\rm ex}/m_{\rm ex}$ we obtain the standard 
quantum-optics results provided 
\be
\sqrt{C^4 +\big\langle 1/n\big\rangle (C^2-C^4)}\approx C^2.
\ee 
For $N=3$
\be
\langle \bbox 0|\bbox 1_\lambda^3|\bbox 0\rangle
&=&
\sum_{n=1}^\infty
\frac{1}{n^3}\Big(
n C^2+3n(n-1)C^4+(n^3-3n^2+2n)C^6\Big)p_n\nonumber\\
&=&
C^6+
3C^4\big(1-C^2\big)\big\langle 1/n\big\rangle +
C^2\big(1-3C^2+2C^4\big)\big\langle 1/n^2\big\rangle.
\ee
As before the result becomes the standard one if the approximation
\be
\sqrt{
C^6+
3C^4\big(1-C^2\big)\big\langle 1/n\big\rangle +
C^2\big(1-3C^2+2C^4\big)\big\langle 1/n^2\big\rangle
}\approx C^3
\ee
is justified. Let us note that in the standard cananonical quantum optics 
one considers vacuum consisting of an {\it infinite\/} number of oscillators 
and such subtleties are trivially ignored. 

\subsection{Spontaneous emission of two different photons}

By the same argument as before the proportionality factor we need to estimate
in second-order perturbation theory is 
\be
\frac{\langle \bbox 1_\lambda,\bbox 1_{\lambda'}|
\bbox a_{\lambda}^{\dag }\bbox a_{\lambda'}^{\dag }|\bbox 0\rangle}
{\langle \hat 1_\lambda,\hat 1_{\lambda'}|
\hat a_{\lambda}^{\dag }\hat a_{\lambda'}^{\dag }|\hat 0\rangle}
&=&
\sqrt{
\langle \bbox 0|
\bbox 1_{\lambda}
\bbox 1_{\lambda'}
|\bbox 0\rangle
}
=
\sqrt{
1- \big\langle 1/n\big\rangle}C^2\approx C^2
\ee
so the result agrees with the canonical one. It is also in perfect agreement 
with the explicit calculations given in \cite{MCv2}. 

\subsection{Stimulated emission}

The last example we will discuss is the first order calculation 
of the transition amplitude 
$|\bbox N_\lambda,A\rangle \to |(\bbox {N+1})_\lambda,B\rangle$.
The appropriate proportionality coeeficient is
\be
\sqrt{
\frac{\langle \bbox 0|
\bbox 1_{\lambda}^{N+1}
|\bbox 0\rangle}{
\langle \bbox 0|
\bbox 1_{\lambda}^{N}
|\bbox 0\rangle}}\approx C
\ee
provided the values of $\big\langle 1/n\big\rangle$, 
$\big\langle 1/n^2\big\rangle$, etc. are sufficiently small.
This is the correct result since $C$ gets absorbed into the renormalized
coupling constant.

\section{Blackbody radiation}

The final test of the new formalism we want to perform is the problem of 
blackbody radiation. Planck's famous formula \cite{Planck}
\be
\varrho(\omega)
=\frac{\hbar}{\pi^2c^3}\frac{\omega^3}{e^{\beta\hbar\omega}-1}
=\frac{\hbar}{\pi^2c^3}\omega^3 \overline{N}_\omega,
\ee
where $\overline{N}_\omega$ is the average number of excitations of
an oscillator in inverse temperature $\beta$,
is one of the first great sucesses of quantum radiation theory and
marks the beginning of quantum mechanics. 
Contemporary measurements of $\varrho(\omega)$ \cite{COBE,COBE99} 
performed by means of COBE (Cosmic Background Explorer) are in a very
good agreement with the Planck law. The data have been carefully
analyzed in the context of nonextensive statistics
\cite{Tsallis1,Tsallis2} in search of possible deviations from
extensivity. The result that comes out systematically is
$|q-1|<10^{-4}$ where $q$ is the Tsallis parameter. The case $q=1$
corresponds to the exact Planck formula. If 
there are any corrections whatever, they must be quite
small. 

The derivation of the formula given by Einstein is based on the properties 
of spontaneous and stimulated emissions. As we have seen above there may occur
differences with respect to the standard formalism but under reasonable 
assumptions they may be expected to be small. 

Below we follow another standard route which consists basically of two
steps. First, one counts the number of different wave vectors $\vec
k$ such that $c|\vec k|\in [\omega,\omega +\Delta\omega]$. Second,
one associates with each such a vector an oscillator and counts the
average number of its excitations assuming the Boltzmann-Gibbs
probability distribution at temperature $T$ and chemical potential
$\mu=0$. The latter assumption is justified by the fact that the
number of excitations of the electromagnetic field is not conserved in
atom-light interactions. 

In the new model the situation is slightly different since there
exists an additional conserved quantum number: The number of {\it
oscillators\/}. As we have seen in previous calculations the
Hamiltonian is block-diagonal with respect to $\oplus$ but changes
the number of excitations in each $N$-oscillator subspace of the
direct sum. 
The state vectors at the multi-oscillator level are symmetric with
respect to permutations of the oscillators and therefore the
oscillators themselves have to be
regarded as bosons whose number is conserved and their chemical potential is
$\mu\neq 0$. However, their excitations should be regarded as bosons
with vanishing chemical potential. 

The eigenvalues of $\cal H$
\be
E_{m,n}=m\hbar\omega\Big(n+\frac{1}{2}\Big).
\ee
corresponding to the oscillator whose frequency
is $\omega$ are parametrized by two natural numbers: $m$ (the number
of oscillators) and $n$ (the number of excitations). Assuming the
standard Boltzmann-Gibbs statistics we obtain the probabilities 
\be
p_{m,n}=Z^{-1}e^{-\beta [m\hbar\omega(n+\frac{1}{2})-m\mu]}
\ee
where 
\be
Z &=& 
\sum_{m=1}^\infty e^{\beta m(\mu+\hbar\omega/2)}
\frac{e^{-\beta m\hbar\omega}}
{1-e^{-\beta m\hbar\omega}}.\label{Z}
\ee
The Lambert series \cite{F} 
\be
\sum_{m=1}^\infty a_m\frac{x^m}{1-x^m}\label{Lambert}
\ee
is convergent for any $x$ if $\sum_{m=1}^\infty a_m$ is convergent. 
Otherwise (\ref{Lambert}) converges for exactly those $x$ for
which the power series $\sum_{m=1}^\infty a_mx^m$ does.
In (\ref{Z}) $a_m=e^{\beta m(\mu+\hbar\omega/2)}$ and 
$\sum_{m=1}^\infty a_m<\infty$ if $\mu+\hbar\omega/2<0$. 
If $\mu+\hbar\omega/2\geq 0$ we still have convergence of (\ref{Z}) 
as long as 
$\sum_{m=1}^\infty
e^{-\beta m[\frac{1}{2}\hbar\omega -\mu]}<\infty$. The upper limit
imposed on $\mu$ by the finiteness of $Z$ is therefore 
$\mu<\frac{1}{2}\hbar\omega$. In what follows we assume that 
$\mu$ is 
$\omega$-independent and therefore $\mu\leq 0$.

The appropriate average number of excitations  is
\be
\overline{n}_\omega
&=&
Z^{-1}\sum_{m=1}^\infty\sum_{n=0}^\infty mn 
e^{-\beta [m\hbar\omega(n+\frac{1}{2})-m\mu]}
\ee
and the Planck formula is replaced by
\be
\varrho_{\rm new}(\omega)
&=&
\frac{\hbar}{\pi^2c^3}\omega^3\overline{n}_\omega.
\ee
It is easy to show that $\varrho_{\rm new}(\omega)$ tends to the Planck 
distribution with $\mu\to -\infty$. To see this consider a more general 
series
\be
Z^{-1}\sum_{m=1}^\infty\sum_{n=0}^\infty mn 
q_m e^{-\beta m\hbar\omega(n+\frac{1}{2})}\label{series}
\ee
where $Z$ is the normalization factor and 
$\sum_{m=1}^\infty q_m<\infty$. If $q_1=1$ and $q_m=0$ for $m>1$ 
then (\ref{series}) is just the exact Planckian formula. 
Factoring out $e^{-\beta|\mu|}$ in both the numerator and the denominator 
of $\overline{n}_\omega$ we obtain $q_1=1$ and $q_m=e^{-\beta|\mu|(m-1)}$ 
for $m>1$. For $|\mu|\to \infty$ all $q_m$, for $m>1$, vanish 
and the limiting distribution is Planckian. 

This proves that an experimental agreement with the ordinary Planck's
$\varrho(\omega)$ cannot rule out our modification but can, at most, 
set a lower bound on an admissible value of $|\mu|$.
However, assuming that $\mu$ has 
some finite and fixed value it should be in principle measurable. 
The plots show that the modifications become visible around
$\mu\approx -3k_BT$. Assuming that the chemical potential is
temperature independent, say $\mu=-k_BT_0$, we obtain a kind of
critical temperature $T_{\rm critical}\approx T_0/3$ above which the
ratio $\mu/(k_BT)$ is small enough to make the modifications of the
distribution observable. For $T<T_{\rm critical}$ the distribution
should be given by the Planck law; for $T>T_{\rm critical}$ the
distribution should approach the $\mu=0$ distribution, i.e. this
would be a Planck-type curve but with the maximum lowered and shifted
towards higher energies.

Fig.~1 shows the plots of $\varrho_{\rm new}(\omega)$ for $\mu=0$
(lower dotted),
$\mu= -0.8 k_B T$ (upper dotted), and $\mu= -10 k_B T$ (solid). 
The thick dashed curve is the Planck distribution. The curve obtained
for $\mu=-10 k_B T$ is indistinguishable from the Planck
distribution. 
The plot does not change if one takes $\mu< -10
k_B T$ and differences are not visible even if one plots the distributions
in logarithmic scales (not shown here).
This is a numerical proof that the distribution we have
obtained on the basis of the modified quantization tends {\it very quickly\/}
to the Planck one as $\mu\to -\infty$. 
It is instructive to compare the modification we have predicted with
those arising from nonextensive statistics. 
The two thin dashed lines represent Tsallis distributions resulting
from nonextensive formalism for $q=0.95$ (lower) and $q=1.05$ (upper). 
The modifications we have derived are therefore qualitatively different from
those resulting from Tsallis statistics. 
\begin{figure}
%\begin{center}
\epsfxsize=4in \epsfbox{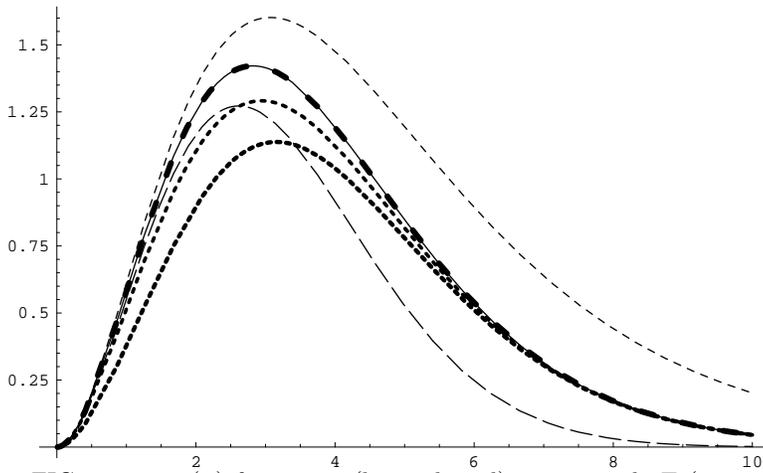}
\caption{
$\varrho_{\rm new}(\omega)$ for $\mu=0$
(lower dotted),
$\mu= -0.8 k_B T$ (upper dotted), and $\mu= -10 k_B T$ (solid). The 
energy range is $0.01 k_B T<\hbar\omega <10 k_B T$.
The thick dashed curve is the Planck distribution. The curve obtained
for $\mu=-10 k_B T$ is indistinguishable from the Planck
distribution.
The two thin dashed lines represent Tsallis distributions resulting
from the Tsallis formalism for $q=0.95$ (lower) and $q=1.05$ (upper).
Since $\varrho_{\rm new}<\varrho$ at least in the neighborhood of the
maximum, the new distribution has to be compared with $q<1$
statistics. The curves are qualitatively different. In particular,
all $q<1$ distributions require an energy cut-off which does not
occur for $\varrho_{\rm new}$.}
%\end{center}
\end{figure}
\begin{figure}
%\begin{center}
\epsfxsize=4in \epsfbox{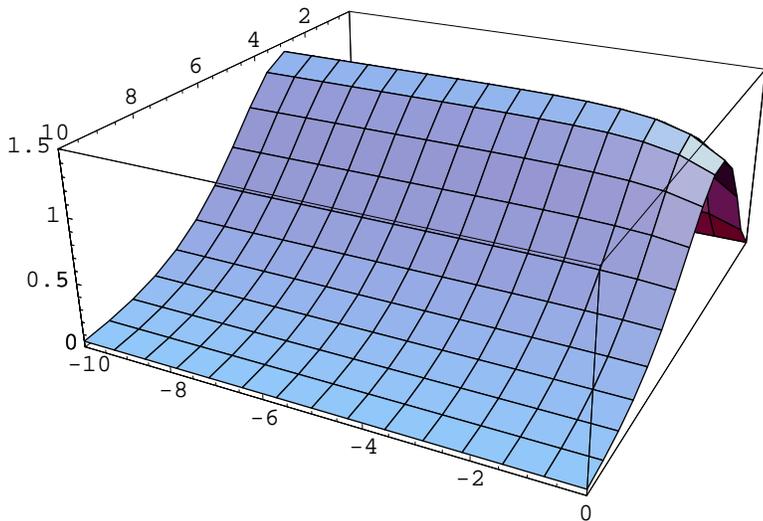}
\caption{
$\varrho_{\rm new}(\omega)$ for $-10\leq -T_0/T\leq 0$. The cut
through $T_0/T=10$ is practically indistinguishable from the Planck
distribution.}
%\end{center}
\end{figure}
\begin{figure}
%\begin{center}
\epsfxsize=4in \epsfbox{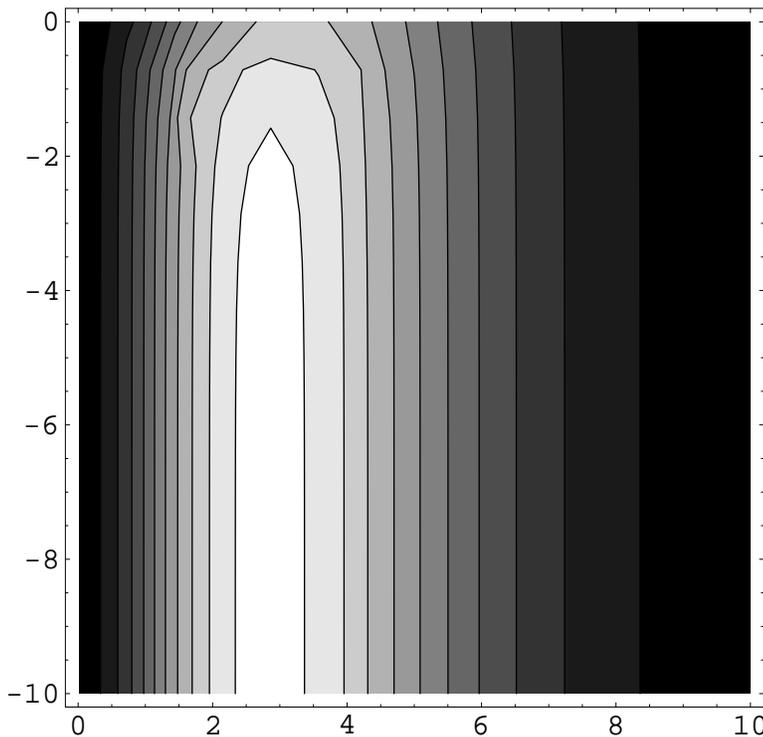}
\caption{
Contour plot of $\varrho_{\rm new}(\omega)$ for $-10\leq -T_0/T\leq
0$. The fast convergence to Planckian $\varrho(\omega)$ (as $T_0/T$ increases)
is clearly seen. 
}
%\end{center}
\end{figure}

\section{Conclusions}

``A theory that is as spectacularly successful as quantum
electrodynamics has to be more or less correct, although we may not
be formulating it in just the right way" \cite{Dreams}. The above
quotation from Weinberg could serve as a motto opening our paper. The
main idea we have tried to advocate was that the standard canonical
quantization procedure is, in certain sense, {\it too classical\/} to
be good. 

The reasons for such a choice of quantization  
could be both historical and sociological and may be
rooted in the fact that the idea of quantizing the field was
formulated before the real development of modern quantum mechanics. 
In oscillations of a simple pendulum it may be justified to treat
$\omega$ as an external parameter defining the system (via, say, the
length of the pendulum). But oscillations of the electromagnetic
field do not seem to have such a ``mechanical" origin 
and it is more natural to think of the
spectrum of frequencies as eigenvalues of some Hamiltonian. That is
exactly what happens with other quantum wave equations. 

We have defined the quantum electromagnetic field as an oscillator
that can exist in a superposition of different frequencies (or,
rather, wave vectors). This should not be confused with the classical
superpositions of frequencies created by, say, a guitar string. The
superpositions we have in mind dissappear at the classical level. 

Once one accepts this viewpoint it becomes 
clear how to quantize the
field at the level of a single oscillator. We do not need many
oscillators to perform the field quantization. But 
there is no reason to believe that all the possible fields can be
described by the same single oscillator. And even more: We know that
the structure of the one-oscillator Hilbert space is not rich enough to
describe multi-particle entangled states and there is no doubt that
such states are physical. The next step, performed already {\it after\/}
the quantization, is to consider fields consisting of 1, 2, 3 and more
oscillators, and even existing in superpositions of different numbers
of them. The resulting structure is in many respects analogous to the Fock
space so that the procedure can be 
(although somewhat misleadingly) referred to as ``second quantization''.  
What is essential we do not need the vacuum state understood as
the {\it unique\/} cyclic vector of the GNS construction. 

On the other hand, there exist vacuum {\it states\/}. These are all
the states describing ground states of the oscillators. 
They correspond to concrete {\it finite\/} average values of energy. A
general vacuum state is therefore a superposition of different
eigenstates of a free Hamiltonian and is not, in itself, an
eigenstate of the Hamiltonian. 

One technical assumption we have made is the resolution of identity property of
the non-CCR algebra. This assumption is clearly satisfied at the one-oscillator
level. The remaining assumptions are standard. 
The system is described by laws of
ordinary quantum mechanics so that to compute concrete problems we
can use standard methods. Perturbation theory
leads to structures we know from the standard Feynmann diagrams.
The blackbody radiation is calculated by means of the standard
Boltzmann-Gibbs statistics. 

There are still several unexplored possibilities. To give an example, 
we have seen in explicit calculations that differences between the canonical
and non-canonical formalisms consist of the factors occuring in 
perturbative expansions of amplitudes. These factors explicitly depend on
the choice of the RHS of the non-CCR algebra. As such, they point into
possible experiments testing {\it directly\/} the algebra of canonical
commutation relations. The meaning of such tests is, at the present stage, 
obscured by the lack of proper understanding of the role of renormalisation.
Actually, one should not expect here precise results since even the 
nonrelativistic canonical quantum electrodynamics is non-renormalizable. 
The simple illustration we have used, namely the one with the cut-off and 
flat vacuum, shows that the new theory is not that far from the canonical 
one as one might expect. 

Let us close these remarks with another quotation: ``Present quantum 
electrodynamics contains many very important `elements of truth', but also 
some clear `elements of nonsense'. Because of the divergences and ambiguities, 
there is general agreement that a rather deep modification of the theory 
is needed, but in some forty years of theoretical work, nobody has seen how to 
disentengle the truth from the nonsense. In such a situation, one needs 
more experimental evidence, but during that same forty years we have found 
no clues from the laboratory as to what specific features of QED might be
modified. Even worse, in the absence of any alternative theory whose 
predictions differ from those of QED in known ways, we have no criterion 
telling us {\it which\/} experiments would be relevant ones to try. 

It seems useful, then, to examine the various disturbing features of QED, 
which give rise to mathematical or conceptual difficulties, to ask
whether present empirical evidence demands their presence, and to explore 
the consequences of the modified (although perhaps rather crude and 
incomplete) theories in which these features are removed. Any difference 
between the predictions of QED and some alternative theory, corresponds 
to an experiment which might distinguish between them; if it appears untried
but feasible, then we have the opportunity to subject QED to a new test 
in which we know just what to look for, and which we would be very unlikely 
to think of without the alternative theory. For this purpose, the alternative
theory need not be worked out as completely as QED; it is sufficient 
if we know in what way their predictions will differ in the area of 
interest. Nor does the alternative theory need to be free of defects in 
all other respects; for if experiment should show that it contains just a 
single `element of truth' that is {\it not\/} in QED, then the alternative 
theory will have served its purpose; we would have the long-missing clue
showing in what way QED must be modified, and electrodynamics (and, I suspect,
much more of theoretical physics along with it) could get moving again
`` \cite{Janes}. 
\acknowledgements

This work was done partly during my stays in Antwerp University, UIA, and 
Arnold Sommerfeld
Institute in Clausthal. I gratefully acknowledge a support from the
Alexander von Humboldt Foundation and the Polish-Flemish grant No. 007. 
I'm indebted to Prof. Iwo
Bia{\l}ynicki-Birula, Robert~Alicki, Jan Naudts, Maciej Kuna, 
and Wolfgang Luecke 
for critical comments, and Pawe{\l} Syty for a stimulating discussion 
on small $\omega$'s. I express my gratitude  
to prof. Heinz-Dietrich Doebner for various help.


\begin{references}
\bibitem{BHJ}M. Born, W. Heisenberg, and P. Jordan, Z. Phys. {\bf
35}, 557 (1925)
\bibitem{Sch}E. Schr\"odinger, Ann. Phys. {\bf 79}, 361 (1926); {\it ibid.\/} 
{\bf 79}, 489 (1926).
\bibitem{Jammer}M. Jammer, {\it The Conceptual Development of 
Quantum Mechanics\/} (Mc Graw-Hill, New York, 1966).
\bibitem{Janes}E. T. Janes, in {\it Coherence and Quantum Optics III\/}, 
edited by L. Mandel and E. Wolf (Plenum, New York, 1973). 
\bibitem{MS}A. Casado, T. W. Marshall, R. Risco Delgado, and E. Santos, Phys.
Rev. A {\bf 55}, 3879 (1997); {\it ibid.\/} Phys.
Rev. A {\bf 56}, 2477 (1997). 
\bibitem{D}P. A. M. Dirac, Proc. Roy. Soc. A {\bf 112}, 661 (1926); 
ibid. {\bf 114}, 243 (1927).
\bibitem{Lambda}J. D. Cohn, astro-ph/9807128.
\bibitem{Busch}P. Busch, M. Grabowski, P. J. Lahti, {\it Operational
Quantum Physics\/} (Springer, Berlin, 1995).
\bibitem{MCv2}M. Czachor, quant-ph/0002003v2.
\bibitem{Dreams}S. Weinberg, {\it Dreams of a final theory\/}
(Vintage, New York, 1994). 
%\bibitem{C-T}C. Cohen-Tannoudji, J. Dupont-Roc, and G. Grynberg, {\it
%Atom-photon interactions: Basic processes and applications\/} (Wiley,
%New York, 1992).
\bibitem{Planck}M. Planck, Verh. Deut. Phys. Gessellsch. {\bf 2}, 237
(1900).
\bibitem{COBE}
L. Page and D. Wilkinson, Rev. Mod. Phys. {\bf 71}, S173 (1999). 
\bibitem{COBE99}G. F. Smoot, astro-ph/9902027.
\bibitem{Tsallis1}C. Tsallis, F. C. Sa Bareto, and E. D. Loh, Phys.
Rev. E {\bf 52}, 1447 (1995).
\bibitem{Tsallis2}A. R. Plastino, A. Plastino, and H. Vucetich, Phys.
Lett. A {\bf 207}, 42 (1995).
\bibitem{F}G. M. Fichtenholz, {\it Course of Differential and
Integral Calculus\/}, vol. 2 (Fizmatgiz, Moscow, 1959) (in Russian). 
\end{references}
\end{document}